\documentclass[12pt,preprint]{aastex}

\slugcomment{To appear in March 1, 2007 ApJ v657 n1}

\shorttitle{Surface Gravity of $\eta$ Boo}

\shortauthors{van Belle, Ciardi \& Boden}

\begin{document}


\title{Measurement of the Surface Gravity of $\eta$ Boo}


\author{Gerard T. van Belle\altaffilmark{1}, David R. Ciardi, Andrew F. Boden}
\affil{Michelson Science Center, California Institute of
Technology, Pasadena, CA 91125\\
gerard, bode, ciardi@ipac.caltech.edu}


\altaffiltext{1}{For preprints, please email
gerard@ipac.caltech.edu.}


\begin{abstract}

Direct angular size measurements of the G0IV subgiant $\eta$ Boo
from the Palomar Testbed Interferometer are presented, with
limb-darkened angular size of $\theta_{LD}=
2.1894^{+0.0055}_{-0.0140} $ mas, which indicate a linear radius of
$R=2.672 \pm 0.028 R_\odot$. A bolometric flux estimate of $F_{BOL}
=  22.1 \pm 0.28\times 10^{-7}$ erg cm$^{-2}$s$^{-1}$ is computed,
which indicates an effective temperature of $T_{EFF}=6100 \pm 28$ K
and luminosity of $L = 8.89 \pm 0.16 L_\odot$ for this object.
Similar data are established for a check star, HD 121860. The $\eta$
Boo results are compared to, and confirm, similar parameters
established by the {\it MOST} asteroseismology satellite. In
conjunction with the mass estimate from the {\it MOST}
investigation, a surface gravity of $\log g=3.817 \pm 0.016$ [cm
s$^{-2}$] is established for $\eta$ Boo.

\end{abstract}



\section{Introduction}

The bright G0 IV star $\eta$ Bo\"otis (8 Boo, HR 5235, HD 121370) is
an interesting target to study given its place on the H-R diagram,
and its implications upon stellar modeling.  Solar-like oscillations
were detected for $\eta$ Boo by
\citet{1995AJ....109.1313K,2003AJ....126.1483K}. $\eta$ Boo was
recently observed with the Microvariability and Oscillations of
STars ({\it MOST}) satellite \citep{2005ApJ...635..547G}, a 15-cm
aperture satellite observatory orbited in 2003 June by the Canadian
Space Agency \citep{2003PASP..115.1023W}. Given that the convective
envelope of $\eta$ Boo is expected to be very thin, containing less
than 1\% of the total mass of the star, observations by {\it MOST}
were motivated by the possibility of detecting $g$-modes, along with
the $p$-modes indicative of turbulent convection. In detecting eight
consecutive low-frequency radial $p$-modes for this G0IV star, the
{\it MOST} team was able to estimate many stellar parameters,
including temperature, age, and mass.  Additionally, new models
examined by \citet{2006ApJ...636.1078S} are able to match
simultaneously the space- and ground-based pulsation data for $\eta$
Boo through the inclusion of turbulence in the stellar models.

Using interferometry to obtain a {\it direct}, absolute measurement
of this object's linear size and effective temperature, in
conjunction with the the mass estimate from the models fit to the
{\it MOST} data, we may also infer its surface gravity,
$\log g$.  Values of $\log g$ are frequently utilized in stellar
modeling and spectroscopy, and a direct characterization of $\log g$
for this slightly evolved object is of significant utility.
We
show that the combination of the high-precision photometry of {\it
MOST} with the high-spatial resolution observations of
the Palomar Testbed Interferometer ({\it PTI})
make for a potent combination for uniquely interpreting a star's
astrophysical parameters.

\section{Interferometric Observations and Data Reduction}\label{sec_PTIData}

{\it PTI} is a
three-element long-baseline interferometer, which combines two of
its three 40-cm apertures to make measurements of fringe visibility,
$V^2$, for astronomical sources. These measurements are made at
either $H-$ (1.6 $\mu$m) or $K-$band (2.2 $\mu$m)  with PTI; for this
particular investigation, the K band was employed, being spectrally
dispersed into 5 `narrow' bands across K, centered at 2.009, 2.106,
2.203, 2.299 and 2.396 $\mu$m.
 For all of these observations,
{\it PTI}'s 109.8-m N-S baseline was utilized;
details on {\it PTI} can be found in
\citet{1999ApJ...510..505C}.

\begin{deluxetable}{lcccccccc}
\tablecolumns{9}
\tablewidth{0pc}
\tablecaption{Results from spectral energy distribution fitting,
including reddening, estimated angular size, and bolometric flux at source.\label{table0}}
\tablehead{
          &         \colhead{RA} &         \colhead{DE} &
                    \colhead{$V$} &          \colhead{$K$} &
                     \colhead{Spectral} &         \colhead{$A_V$} &
                           \colhead{$\theta_{EST}$} &       \colhead{$F_{BOL}$} \\
          &         \colhead{(J2000)} &         \colhead{(J2000)} &
                    \colhead{(mag)} &          \colhead{(mag)} &
                     \colhead{Template} &         \colhead{(mag)} &
                           \colhead{(mas)} &       \colhead{(erg cm$^{-2}$s$^{-1}$)}
}
\startdata


$\eta$ Boo:\\
  HD121370 & 13 54 41.002 & +18 23 55.30 &       2.68 &      1.485 &       G0IV &       $ 0.000 \pm 0.011 $ & $ 2.320 \pm 0.072 $ & $ (221 \pm 2.8)\times 10^{-8} $ \\

\multicolumn{5}{l}{Resolved check star:}\\

  HD121860 & 13 58 01.602 & +07 27 48.40 &        7.5 &      2.448 &       M2III &        $ 0.927 \pm 0.043 $ & $ 2.010 \pm 0.109 $ & $ (25.4\pm 1.3)\times 10^{-8} $ \\

\multicolumn{5}{l}{Calibrators:}\\

  HD117176 & 13 28 25.809 & +13 46 43.63 &       5.000 &        3.500 &        G5V & $ 0.117 \pm 0.015 $ & $ 0.986 \pm 0.020 $ & $ (31.5 \pm 0.6)\times 10^{-8} $ \\
  HD120136 & 13 47 15.743 & +17 27 24.86 &        4.5 &      3.507 &       F5IV &      $ 0.214 \pm 0.017 $ & $ 0.892 \pm 0.042 $ & $ (49.2\pm 1.2)\times 10^{-8} $        \\
  HD121107 & 13 53 12.931 & +17 55 58.33 &      5.711 &      4.077 &      G5III & $ 0.002 \pm 0.043 $ & $ 0.837 \pm 0.052 $ & $ (16.6\pm 0.8)\times 10^{-8} $ \\
  HD121560 & 13 55 49.994 & +14 03 23.41 &        6.1 &      4.843 &        F6V &        $ 0.105 \pm 0.016 $ & $ 0.442 \pm 0.010 $ &        $ (10.1 \pm 0.1)\times 10^{-8} $ \\

\enddata
\end{deluxetable}



$\eta$ Boo was observed along with the unresolved calibration
sources HD117176 (70 Vir), HD120136 ($\tau$ Boo), HD121107,
HD121560, on 18 nights between 2000 and 2005. In addition to $\eta$
Boo, a resolved check star, HD121860, was observed as a means to
monitor system performance. All of the calibrators did not exceed
{\it PTI}'s point-like calibrator angular size criterion of
$\theta_{EST}<1.0$ mas \citep{2005PASP..117.1263V} for absolute
measurement calibration. A previous interferometric measure of
$\eta$ Boo's size was made by \citet{2005A&A...436..253T}, but
resolved calibration sources used in this study (due to sensitivity
limitations) makes the resulting $\eta$ Boo diameter estimate
subject to potential systematic errors. Two of the four calibrators,
70 Vir and $\tau$ Boo, are associated with radial velocity planets
\citep{1996ApJ...464L.147M,1997ApJ...474L.115B}, but no evidence has
been found for $V^2$ variations indicative of face-on binary stars
that could supplant the RV planet interpretation for these objects.
The relevant observing parameters are found in Table \ref{table0},
along with parameters to be derived in \S \ref{sec_SED_fitting}.

\begin{figure*}
\plotone{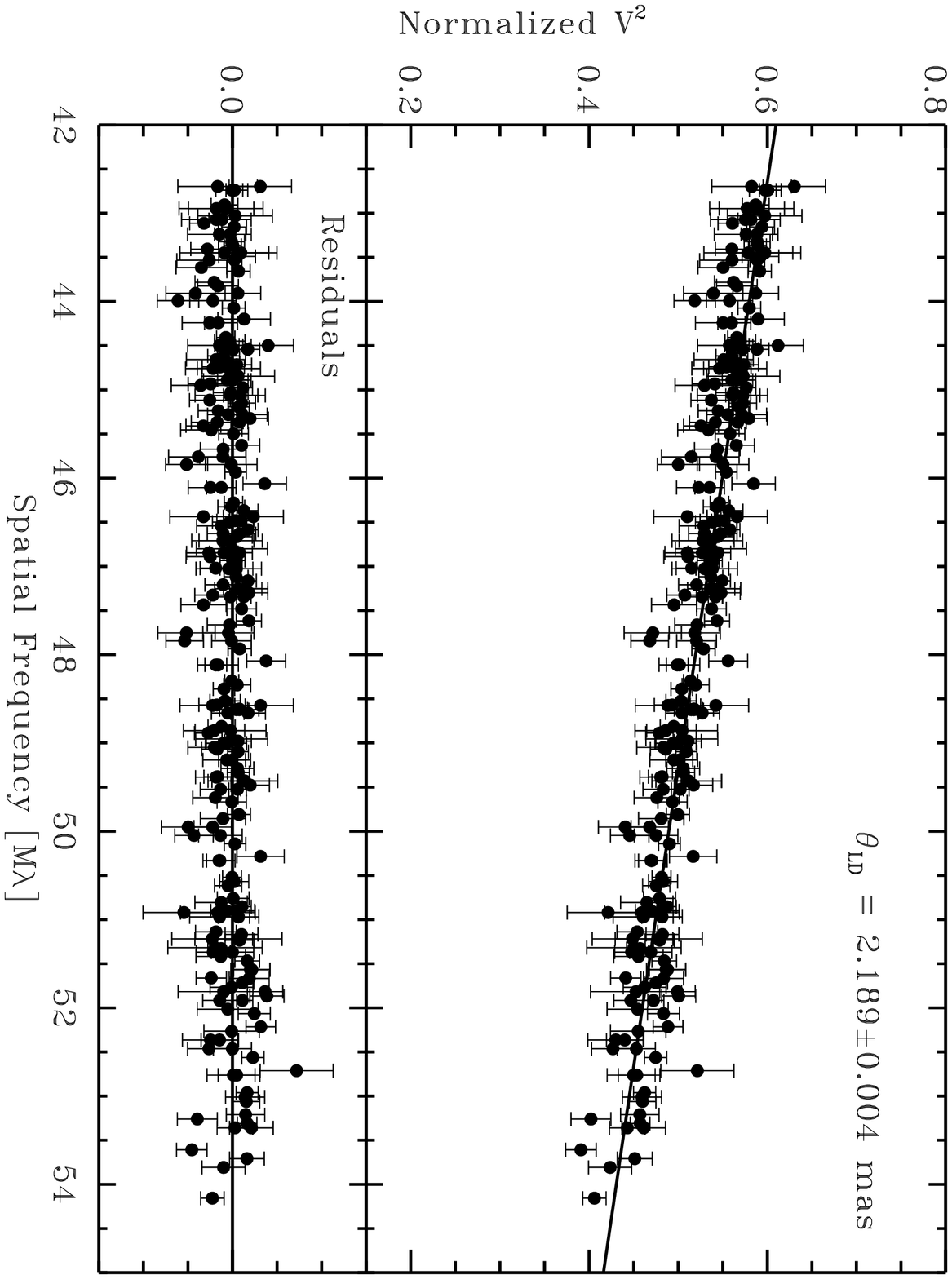} \caption{\label{fig_etaBooAbsV2} Absolute
visibility data for $\eta$ Boo, as discussed in \S
\ref{sec_PTIData}. The line fit to the data is the visibility
function corresponding to a $2.1894 \pm 0.0038$ mas limb-darkened
angular disk diameter.  See \S 3.2 and 4 for a discussion of our
final uncertainty estimate for the $\eta$ Boo angular size.}
\end{figure*}

The calibration of the $\eta$ Boo $V^2$ data is performed by
estimating the interferometer system visibility ($V^2_{\textrm{\tiny
SYS}}$) using the calibration source with a model angular diameter
and then normalizing the raw $\eta$ Boo visibility by
$V^2_{\textrm{\tiny SYS}}$ to estimate the $V^2$ measured by an
ideal interferometer at that epoch
\citep{1991AJ....101.2207M,1998SPIE.3350..872B,2005PASP..117.1263V}.
Uncertainties in the system visibility and the calibrated target
visibility are inferred from internal scatter among the data in an
observation using standard error-propagation calculations
\citep{1999PASP..111..111C}. Calibrating our point-like calibration
objects against each other produced no evidence of systematics, with
all objects delivering reduced $V^2 = 1$. The observation dates,
calibrated visibilities for each wavelength bin, $(u,v)$
coordinates, and observation hour angle are presented in Table
\ref{table1} for $\eta$ Boo and Table \ref{table1b} for HD121860.
Plots of the absolute visibility data for $\eta$ Boo are found in
Figure \ref{fig_etaBooAbsV2}. {\it PTI}'s limiting night-to-night
measurement error is $\sigma_{V^2_{\textrm{\tiny SYS}}}\approx 1.5
-1.8$\%, the source of which is most likely a combination of
effects: uncharacterized atmospheric seeing (in particular,
scintillation), detector noise, and other instrumental effects. This
measurement error limit is an empirically established floor from the
previous study of \citet{bod99}.

\begin{deluxetable}{rrrrrrrrrr}
\tablecolumns{8}
\tabletypesize{\footnotesize}
\tablewidth{0pc}
\tablecaption{Calibrated visibility data obtained by PTI for $\eta$ Boo.\label{table1}}
\tablehead{
\colhead{Date} &\colhead{UT}
      & \colhead{$ V^2 (2.009 \mu$m)}
      & \colhead{$ V^2 (2.106 \mu$m)}
      & \colhead{$ V^2 (2.203 \mu$m)}
      & \colhead{$ V^2 (2.299 \mu$m)}
      & \colhead{$ V^2 (2.396 \mu$m)}
&\colhead{$U$ (m)} &\colhead{$V$ (m)} &\colhead{HA}
} \startdata

  4/9/2000 &       7.73 & $ 0.391 \pm 0.017 $ & $ 0.529 \pm 0.033 $ & $ 0.479 \pm 0.015 $ & $ 0.510 \pm 0.025 $ & $ 0.454 \pm 0.023 $ &     -48.46 &     -96.13 &      -0.78 \\

  4/9/2000 &       7.98 & $ 0.402 \pm 0.022 $ & $ 0.551 \pm 0.033 $ & $ 0.493 \pm 0.020 $ & $ 0.529 \pm 0.028 $ & $ 0.465 \pm 0.029 $ &     -44.96 &     -97.12 &      -0.53 \\

  4/9/2000 &       9.14 & $ 0.579 \pm 0.050 $ & $ 0.556 \pm 0.023 $ & $ 0.537 \pm 0.026 $ & $ 0.510 \pm 0.039 $ & $ 0.499 \pm 0.020 $ &     -26.82 &    -100.58 &       0.63 \\

 4/28/2000 &       5.93 & $ 0.526 \pm 0.020 $ & $ 0.441 \pm 0.017 $ & $ 0.507 \pm 0.020 $ & $ 0.406 \pm 0.013 $ & $ 0.480 \pm 0.023 $ &     -55.30 &     -93.75 &      -1.34 \\

  5/2/2000 &       6.44 & $ 0.566 \pm 0.011 $ & $ 0.518 \pm 0.009 $ & $ 0.487 \pm 0.011 $ & $ 0.457 \pm 0.011 $ & $ 0.558 \pm 0.011 $ &     -45.54 &     -96.96 &      -0.57 \\

  5/2/2000 &       6.70 & $ 0.566 \pm 0.010 $ & $ 0.514 \pm 0.008 $ & $ 0.481 \pm 0.011 $ & $ 0.546 \pm 0.010 $ & $ 0.462 \pm 0.012 $ &     -41.77 &     -97.91 &      -0.31 \\

  5/2/2000 &       7.53 & $ 0.589 \pm 0.008 $ & $ 0.502 \pm 0.010 $ & $ 0.567 \pm 0.010 $ & $ 0.472 \pm 0.012 $ & $ 0.542 \pm 0.008 $ &     -28.50 &    -100.35 &       0.53 \\

  5/2/2000 &       7.71 & $ 0.589 \pm 0.010 $ & $ 0.506 \pm 0.012 $ & $ 0.569 \pm 0.012 $ & $ 0.475 \pm 0.013 $ & $ 0.549 \pm 0.009 $ &     -25.46 &    -100.75 &       0.71 \\

  5/4/2000 &       6.96 & $ 0.558 \pm 0.026 $ & $ 0.475 \pm 0.024 $ & $ 0.550 \pm 0.029 $ & $ 0.521 \pm 0.022 $ & $ 0.453 \pm 0.021 $ &     -35.70 &     -99.18 &       0.09 \\

  5/4/2000 &       8.14 & $ 0.577 \pm 0.042 $ & $ 0.564 \pm 0.035 $ & $ 0.537 \pm 0.036 $ & $ 0.504 \pm 0.040 $ & $ 0.478 \pm 0.049 $ &     -15.66 &    -101.70 &       1.27 \\

 5/14/2000 &       5.85 & $ 0.504 \pm 0.012 $ & $ 0.566 \pm 0.014 $ & $ 0.475 \pm 0.015 $ & $ 0.557 \pm 0.016 $ & $ 0.460 \pm 0.015 $ &     -42.62 &     -97.71 &      -0.36 \\

 5/14/2000 &       7.47 & $ 0.449 \pm 0.045 $ & $ 0.546 \pm 0.031 $ & $ 0.486 \pm 0.034 $ & $ 0.528 \pm 0.035 $ & $ 0.588 \pm 0.042 $ &     -15.85 &    -101.69 &       1.26 \\

 5/30/2000 &       5.58 & $ 0.494 \pm 0.016 $ & $ 0.558 \pm 0.017 $ & $ 0.591 \pm 0.013 $ & $ 0.537 \pm 0.017 $ & $ 0.484 \pm 0.018 $ &     -30.41 &    -100.07 &       0.42 \\

 5/30/2000 &       6.52 & $ 0.495 \pm 0.015 $ & $ 0.587 \pm 0.015 $ & $ 0.574 \pm 0.017 $ & $ 0.544 \pm 0.018 $ & $ 0.482 \pm 0.018 $ &     -14.07 &    -101.81 &       1.36 \\

 5/30/2000 &       7.68 & $ 0.461 \pm 0.034 $ & $ 0.598 \pm 0.018 $ & $ 0.515 \pm 0.014 $ & $ 0.552 \pm 0.014 $ & $ 0.588 \pm 0.013 $ &       7.23 &    -102.14 &       2.52 \\

 6/12/2000 &       3.70 & $ 0.566 \pm 0.016 $ & $ 0.481 \pm 0.020 $ & $ 0.548 \pm 0.018 $ & $ 0.527 \pm 0.019 $ & $ 0.461 \pm 0.024 $ &     -46.17 &     -96.79 &      -0.61 \\

 6/12/2000 &       4.58 & $ 0.566 \pm 0.023 $ & $ 0.544 \pm 0.026 $ & $ 0.481 \pm 0.025 $ & $ 0.521 \pm 0.025 $ & $ 0.455 \pm 0.031 $ &     -32.89 &     -99.68 &       0.27 \\

 6/12/2000 &       5.34 & $ 0.561 \pm 0.016 $ & $ 0.511 \pm 0.027 $ & $ 0.541 \pm 0.025 $ & $ 0.483 \pm 0.030 $ & $ 0.455 \pm 0.028 $ &     -19.93 &    -101.35 &       1.03 \\

 6/12/2000 &       6.11 & $ 0.572 \pm 0.018 $ & $ 0.601 \pm 0.009 $ & $ 0.545 \pm 0.021 $ & $ 0.482 \pm 0.023 $ & $ 0.518 \pm 0.022 $ &      -6.02 &    -102.18 &       1.80 \\

 6/14/2000 &       5.03 & $ 0.588 \pm 0.018 $ & $ 0.501 \pm 0.015 $ & $ 0.538 \pm 0.018 $ & $ 0.573 \pm 0.019 $ & $ 0.488 \pm 0.020 $ &     -23.03 &    -101.03 &       0.85 \\

 3/16/2001 &      10.65 & $ 0.560 \pm 0.036 $ & $ 0.542 \pm 0.029 $ & $ 0.447 \pm 0.019 $ & $ 0.483 \pm 0.023 $ & $ 0.527 \pm 0.023 $ &     -28.07 &    -100.41 &       0.56 \\

 3/16/2001 &      11.20 & $ 0.448 \pm 0.019 $ & $ 0.575 \pm 0.039 $ & $ 0.498 \pm 0.023 $ & $ 0.560 \pm 0.027 $ & $ 0.545 \pm 0.014 $ &     -18.57 &    -101.47 &       1.11 \\

 3/17/2001 &      10.65 & $ 0.453 \pm 0.051 $ & $ 0.597 \pm 0.040 $ & $ 0.571 \pm 0.028 $ & $ 0.512 \pm 0.037 $ & $ 0.545 \pm 0.025 $ &     -27.09 &    -100.54 &       0.62 \\

 3/17/2001 &      11.14 & $ 0.450 \pm 0.053 $ & $ 0.504 \pm 0.040 $ & $ 0.537 \pm 0.040 $ & $ 0.597 \pm 0.042 $ & $ 0.573 \pm 0.041 $ &     -18.51 &    -101.47 &       1.11 \\

 3/17/2001 &      12.08 & $ 0.459 \pm 0.042 $ & $ 0.566 \pm 0.034 $ & $ 0.630 \pm 0.035 $ & $ 0.542 \pm 0.037 $ & $ 0.612 \pm 0.028 $ &      -1.41 &    -102.25 &       2.05 \\

 3/17/2001 &      12.29 & $ 0.421 \pm 0.046 $ & $ 0.510 \pm 0.038 $ & $ 0.488 \pm 0.037 $ & $ 0.582 \pm 0.045 $ & $ 0.557 \pm 0.036 $ &       2.37 &    -102.24 &       2.26 \\

 3/19/2001 &       8.91 & $ 0.537 \pm 0.023 $ & $ 0.515 \pm 0.017 $ & $ 0.486 \pm 0.018 $ & $ 0.457 \pm 0.015 $ & $ 0.424 \pm 0.024 $ &     -51.21 &     -95.25 &      -1.00 \\

 3/19/2001 &       9.76 & $ 0.550 \pm 0.031 $ & $ 0.523 \pm 0.025 $ & $ 0.501 \pm 0.023 $ & $ 0.470 \pm 0.018 $ & $ 0.450 \pm 0.030 $ &     -39.36 &     -98.44 &      -0.15 \\

 3/19/2001 &      10.57 & $ 0.560 \pm 0.019 $ & $ 0.482 \pm 0.015 $ & $ 0.545 \pm 0.023 $ & $ 0.521 \pm 0.020 $ & $ 0.462 \pm 0.024 $ &     -26.11 &    -100.66 &       0.67 \\

 3/22/2001 &       9.88 & $ 0.468 \pm 0.021 $ & $ 0.440 \pm 0.021 $ & $ 0.542 \pm 0.026 $ & $ 0.519 \pm 0.029 $ & $ 0.587 \pm 0.026 $ &     -34.34 &     -99.42 &       0.18 \\

 3/23/2001 &       9.72 & $ 0.519 \pm 0.023 $ & $ 0.468 \pm 0.021 $ & $ 0.500 \pm 0.024 $ & $ 0.446 \pm 0.022 $ & $ 0.427 \pm 0.024 $ &     -35.92 &     -99.14 &       0.08 \\

 3/23/2001 &      10.62 & $ 0.594 \pm 0.014 $ & $ 0.540 \pm 0.006 $ & $ 0.509 \pm 0.012 $ & $ 0.576 \pm 0.011 $ & $ 0.484 \pm 0.014 $ &     -20.83 &    -101.26 &       0.98 \\

 3/24/2001 &       9.57 & $ 0.580 \pm 0.013 $ & $ 0.490 \pm 0.012 $ & $ 0.554 \pm 0.012 $ & $ 0.528 \pm 0.013 $ & $ 0.475 \pm 0.012 $ &     -37.14 &     -98.90 &       0.00 \\

 4/28/2003 &       6.41 & $ 0.562 \pm 0.020 $ & $ 0.535 \pm 0.014 $ & $ 0.511 \pm 0.015 $ & $ 0.479 \pm 0.016 $ & $ 0.451 \pm 0.020 $ &     -50.08 &     -95.62 &      -0.91 \\

 4/28/2003 &       6.70 & $ 0.554 \pm 0.020 $ & $ 0.530 \pm 0.018 $ & $ 0.470 \pm 0.018 $ & $ 0.504 \pm 0.018 $ & $ 0.443 \pm 0.020 $ &     -46.26 &     -96.76 &      -0.62 \\

 4/28/2003 &       7.00 & $ 0.570 \pm 0.017 $ & $ 0.543 \pm 0.015 $ & $ 0.483 \pm 0.016 $ & $ 0.520 \pm 0.015 $ & $ 0.459 \pm 0.022 $ &     -41.97 &     -97.85 &      -0.32 \\

  4/3/2005 &       8.80 & $ 0.590 \pm 0.029 $ & $ 0.517 \pm 0.027 $ & $ 0.585 \pm 0.024 $ & $ 0.556 \pm 0.022 $ & $ 0.522 \pm 0.041 $ &     -38.93 &     -98.52 &      -0.12 \\

  4/3/2005 &       9.19 & $ 0.562 \pm 0.021 $ & $ 0.500 \pm 0.013 $ & $ 0.565 \pm 0.020 $ & $ 0.544 \pm 0.014 $ & $ 0.489 \pm 0.017 $ &     -32.86 &     -99.66 &       0.27 \\

  4/3/2005 &       9.50 & $ 0.592 \pm 0.021 $ & $ 0.548 \pm 0.021 $ & $ 0.579 \pm 0.021 $ & $ 0.517 \pm 0.022 $ & $ 0.501 \pm 0.019 $ &     -27.74 &    -100.44 &       0.58 \\

  4/3/2005 &       9.79 & $ 0.576 \pm 0.036 $ & $ 0.495 \pm 0.027 $ & $ 0.530 \pm 0.037 $ & $ 0.561 \pm 0.039 $ & $ 0.486 \pm 0.022 $ &     -22.63 &    -101.06 &       0.88 \\

  4/5/2005 &       8.30 & $ 0.560 \pm 0.018 $ & $ 0.503 \pm 0.017 $ & $ 0.539 \pm 0.019 $ & $ 0.479 \pm 0.018 $ & $ 0.457 \pm 0.022 $ &     -44.44 &     -97.24 &      -0.49 \\

  4/5/2005 &       8.67 & $ 0.499 \pm 0.012 $ & $ 0.560 \pm 0.015 $ & $ 0.470 \pm 0.015 $ & $ 0.535 \pm 0.016 $ & $ 0.454 \pm 0.021 $ &     -39.01 &     -98.51 &      -0.12 \\

  4/5/2005 &       8.95 & $ 0.471 \pm 0.032 $ & $ 0.441 \pm 0.030 $ & $ 0.539 \pm 0.033 $ & $ 0.515 \pm 0.033 $ & $ 0.430 \pm 0.031 $ &     -34.50 &     -99.38 &       0.17 \\

  4/5/2005 &       9.25 & $ 0.550 \pm 0.028 $ & $ 0.495 \pm 0.025 $ & $ 0.534 \pm 0.034 $ & $ 0.476 \pm 0.025 $ & $ 0.454 \pm 0.034 $ &     -29.68 &    -100.16 &       0.46 \\

  4/5/2005 &       9.55 & $ 0.589 \pm 0.015 $ & $ 0.537 \pm 0.013 $ & $ 0.506 \pm 0.018 $ & $ 0.573 \pm 0.015 $ & $ 0.484 \pm 0.022 $ &     -24.63 &    -100.83 &       0.76 \\

  4/5/2005 &       9.86 & $ 0.582 \pm 0.015 $ & $ 0.494 \pm 0.020 $ & $ 0.527 \pm 0.016 $ & $ 0.567 \pm 0.017 $ & $ 0.469 \pm 0.024 $ &     -19.08 &    -101.41 &       1.08 \\

\enddata
\end{deluxetable}

\begin{deluxetable}{rrrrrrrrrr}
\tablecolumns{8}
\tablewidth{0pc} \tablecaption{Calibrated visibility data obtained
by PTI for HD121860.\label{table1b}} \tablehead{ \colhead{Date}
&\colhead{UT}
      & \colhead{$ V^2 (2.009 \mu$m)}
      & \colhead{$ V^2 (2.106 \mu$m)}
      & \colhead{$ V^2 (2.203 \mu$m)}
      & \colhead{$ V^2 (2.299 \mu$m)}
      & \colhead{$ V^2 (2.396 \mu$m)}
&\colhead{$U$ (m)} &\colhead{$V$ (m)} &\colhead{HA} } \startdata


  4/9/2000 &       7.83 & $ 0.479 \pm 0.026 $ & $ 0.563 \pm 0.035 $ & $ 0.593 \pm 0.017 $ & $ 0.620 \pm 0.037 $ & $ 0.621 \pm 0.046 $ &     -47.91 &     -90.37 &      -0.74 \\

  4/9/2000 &       8.08 & $ 0.524 \pm 0.028 $ & $ 0.568 \pm 0.034 $ & $ 0.592 \pm 0.018 $ & $ 0.605 \pm 0.032 $ & $ 0.620 \pm 0.043 $ &     -44.39 &     -90.77 &      -0.49 \\

  5/2/2000 &       6.53 & $ 0.534 \pm 0.018 $ & $ 0.575 \pm 0.014 $ & $ 0.602 \pm 0.012 $ & $ 0.641 \pm 0.014 $ & $ 0.647 \pm 0.015 $ &     -44.99 &     -90.71 &      -0.53 \\

  5/2/2000 &       6.80 & $ 0.522 \pm 0.018 $ & $ 0.565 \pm 0.014 $ & $ 0.592 \pm 0.008 $ & $ 0.623 \pm 0.013 $ & $ 0.630 \pm 0.012 $ &     -41.11 &     -91.10 &      -0.26 \\

  5/4/2000 &       7.06 & $ 0.556 \pm 0.036 $ & $ 0.585 \pm 0.040 $ & $ 0.621 \pm 0.030 $ & $ 0.643 \pm 0.040 $ & $ 0.640 \pm 0.034 $ &     -35.04 &     -91.61 &       0.13 \\

 6/12/2000 &       3.53 & $ 0.562 \pm 0.040 $ & $ 0.580 \pm 0.030 $ & $ 0.613 \pm 0.032 $ & $ 0.634 \pm 0.029 $ & $ 0.637 \pm 0.026 $ &     -49.23 &     -90.21 &      -0.84 \\

 6/12/2000 &       4.41 & $ 0.570 \pm 0.049 $ & $ 0.597 \pm 0.039 $ & $ 0.626 \pm 0.038 $ & $ 0.646 \pm 0.036 $ & $ 0.651 \pm 0.032 $ &     -36.50 &     -91.50 &       0.04 \\

 3/17/2001 &       9.81 & $ 0.507 \pm 0.078 $ & $ 0.538 \pm 0.081 $ & $ 0.550 \pm 0.082 $ & $ 0.587 \pm 0.072 $ & $ 0.594 \pm 0.063 $ &     -41.43 &     -91.06 &      -0.28 \\

 3/19/2001 &       9.01 & $ 0.506 \pm 0.035 $ & $ 0.542 \pm 0.022 $ & $ 0.569 \pm 0.028 $ & $ 0.595 \pm 0.026 $ & $ 0.612 \pm 0.036 $ &     -50.66 &     -90.01 &      -0.95 \\

 4/28/2003 &       6.51 & $ 0.516 \pm 0.030 $ & $ 0.546 \pm 0.023 $ & $ 0.577 \pm 0.021 $ & $ 0.599 \pm 0.019 $ & $ 0.610 \pm 0.028 $ &     -49.50 &     -90.16 &      -0.86 \\

 4/28/2003 &       6.79 & $ 0.508 \pm 0.035 $ & $ 0.529 \pm 0.030 $ & $ 0.558 \pm 0.026 $ & $ 0.579 \pm 0.024 $ & $ 0.601 \pm 0.032 $ &     -45.74 &     -90.61 &      -0.58 \\

 4/28/2003 &       7.11 & $ 0.521 \pm 0.037 $ & $ 0.559 \pm 0.029 $ & $ 0.596 \pm 0.027 $ & $ 0.617 \pm 0.024 $ & $ 0.633 \pm 0.031 $ &     -41.16 &     -91.08 &      -0.26 \\

\enddata
\end{deluxetable}



\section{Effective Temperature and Radius Determinations}

\subsection{Bolometric Flux Estimates}\label{sec_SED_fitting}

For each of the target and calibrator stars observed in this
investigation, a spectral energy distribution (SED) fit was
performed.  This fit was accomplished using photometry available in
the literature as the input values, with template spectra
appropriate for the spectral types indicated for the stars in
question. The template spectra, from \citet{pic98}, were adjusted to
account for overall flux level and wavelength-dependent reddening,
resulting in an estimate of angular size. Reddening corrections were
based upon the empirical reddening determination described by
\citet{1989ApJ...345..245C}, which differs little from van de
Hulst's theoretical reddening curve number 15 \citep{joh68,dyc96}.
Both narrowband and wideband photometry in the 0.3 $\mu$m to 30
$\mu$m were used as available, including Johnson $UBV$ (see, for
example,
\citet{1963AJ.....68..483E,1972ApJ...175..787E,1971A&A....12..442M}),
Str\"omgren $ubvy\beta$ \citep{1976HelR....1.....P}, Geneva
\citep{1976A&AS...26..275R}, 2Mass $JHK_s$
\citep{2003yCat.2246.....C}, and Vilnius $UPXYZS$
\citep{1972VilOB..34....3Z}; zero-magnitude flux density
calibrations were based upon the values given in \citet{cox00}.


Starting with a reference spectral type and luminosity class as
cited by SIMBAD, template spectra were fit to the photometric data.
Templates in adjacent locations in spectral type and luminosity
class were also tested for best fit, with the fit with best $\chi^2$
being selected in the end for use in this study.  For example, a
star indicated by SIMBAD to be a G0IV would have its photometry fit
to the 9 templates of spectral type F9, G0, and G1, and for
luminosity classes III, IV, and V.  Metallicities for these fits
were assumed to be roughly solar, which is consistent with the values
found for these objects in the references listed in
\citet{1997A&AS..124..299C} and \citet{2001A&A...373..159C} .

\begin{figure*}
\plotone{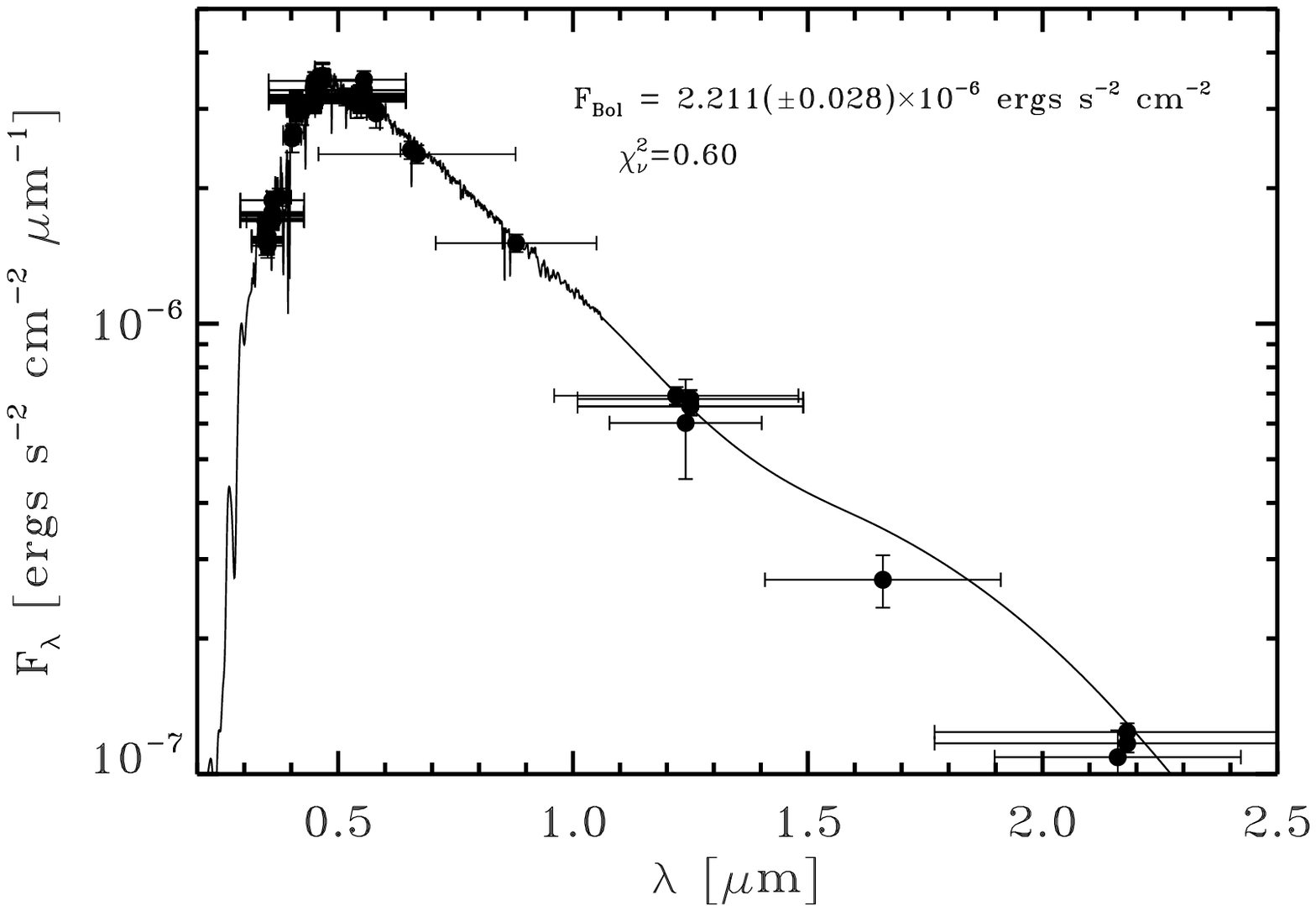}
\caption{\label{fig_HD121370} Spectral energy distribution fitting
for $\eta$ Boo, as discussed in \S \ref{sec_SED_fitting}.
Vertical bars on the data points represent the photometric errors, and
horizontal bars correspond to the bandpass of the photometric filter
being used.}
\end{figure*}

From the best SED fit, estimates were obtained for each star for
their reddening ($A_V$) and bolometric flux ($F_{BOL}$); since
effective temperature was fixed for each of the \citet{pic98}
library spectra, an estimate of angular size ($\theta_{EST}$) was
also obtained. The results of the fitting are given in Table
\ref{table0}. As an example, the SED fitting plot for $\eta$ Boo is
given in Figure \ref{fig_HD121370}.

For our calibration sources, {\it a priori} estimates of their sizes
are necessary to account for residual resolution that may be
afforded by an interferometer's long baselines. With
an expected limb darkened size of $\theta_{EST} \leq 1.00$ mas from
the SED fit, calibrators have predicted $V^2$'s of $>86$\% for a
110-m baseline used at $\lambda=2.2 \mu$m (with $V^2>96$\% expected
for our smallest calibrator, HD 121560). We consider this size
effectively identical to its uniform disk size, since for most of
our potential calibration sources, their effective temperatures are
in excess of $\sim 5000$K. The difference between the uniform
disk and limb darkened sizes is at the few percent level
\citep{dav00,cla03b}, which is far less than our size estimate error
or, in particular, its impact upon the system visibility estimate
derived from observations of our calibrators. A $\leq 5\%$
uncertainty in angular size will contribute, at most, less than
$\leq 1.3\%$ uncertainty to the system visibility
$V^2_{\textrm{\tiny SYS}}$ for {\it PTI}.  The
night-to-night limiting measurement error of {\it PTI} is
$\sigma_{V^2_{\textrm{\tiny SYS}}}\approx 1.5 -1.8$\% \citep{bod99},
any measures of $V^2$ using our calibrators will be free from any
potential bias in its angular size measurement at the $\sigma_\theta
/ \theta \approx 7$\% level for our largest calibrator, HD117176,
and at better levels of insensitivity for our the smaller
calibrators.

\subsection{Angular Sizes of $\eta$ Boo and HD 121860}\label{sec_angularSizes}

We may fit our observed visibility data to a uniform disk approximation to get
an initial estimate of the angular sizes of $\eta$ Boo and HD 121860.  From
the $V^2 = [2 J_1(x) / x]^2$, where $x = \pi B \theta_{UD} \lambda^{-1}$, we get
angular sizes $\theta_{UD}$ of $2.1528 \pm 0.0037$ mas and $2.035 \pm 0.009$ mas for
$\eta$ Boo and HD 121860, respectively, with reduced $\chi^2$ values of 0.90 and 0.48.
Given $\eta$ Boo's low rotation value of 13.5 km/s \citep{2006A&A...446..267R},
rotational oblateness did not need to be considered in the fit \citep{2001ApJ...559.1155V}.

For limb darkened fits, we utilized the visibility function for a limb-darkened stellar
disk as parameterized with a linear limb-darkening coefficient, $\mu_\lambda$
\citep{1974MNRAS.167..475B}:
\begin{equation}
V^2 = \left( {1-\mu_\lambda \over 2} + {\mu_\lambda \over 3}\right)^{-2}
\times
\left[
(1-\mu_\lambda) {J_1 (x) \over x}+\mu_\lambda {j_1 (x) \over x}
\right]^2
\end{equation}
where $x = \pi B \theta_{LD} \lambda^{-1}$.  For these fits we used
the 2.2 $\mu$m coefficients of $\mu_\lambda=0.22$ and 0.38 for
$\eta$ Boo and HD 121860, respectively \citep{1995A&AS..114..247C}.
Examination of the linear limb darkening coefficients from
\citet{1995A&AS..114..247C} indicate that the value of $\mu_\lambda
= 0.22 \pm 0.02$ is sufficient to account for the 5 narrowband
channels of the {\it PTI} data (eg., $\mu_\lambda$ varies by less
than $\Delta \mu_\lambda = 0.04$ between 2.0 $\mu$m and 2.4 $\mu$m).
Fitting our data, we get limb darkened angular sizes of
$\theta_{LD}=2.1894 \pm 0.0038$ mas and $2.100 \pm 0.009$ mas for
$\eta$ Boo and HD 121860, respectively, with no appreciable change
in the reduced $\chi^2$ values as compared to the uniform disk fits.
A previous limb-darkened angular size of $2.25 \pm 0.25$ mas for $\eta$ Boo was measured by \citet{2003AJ....126.2502M}
and is consistent with our measurement.

These errors are sufficiently small that additional sources of error
need to be considered.  First, knowledge of PTI's operational
wavelength has a limit of $\sigma_\lambda \approx 0.01$ $\mu$m; and
Second, the limb darkening coefficient $\mu_\lambda$ is estimated to
be accurate to only $\sigma_\mu \approx 0.02$. Incorporating these
effects into our LD fit for $\eta$ Boo, we find an additional
uncertainty contribution of 0.0040 mas, resulting in $\theta_{LD} =
2.1894 \pm 0.0055$ mas, with no appreciable increase in error for HD
121860, where the measurement error dominates the uncertainty due to
the smaller number of measurements.  We shall return to the estimate
of uncertainty on $\eta$ Boo's angular size in \S
\ref{sec_possibleBinary}, where the limits on our knowledge of
$\eta$ Boo's possible binarity in fact ultimately limits the lower
bound on our knowledge of the star's angular size.

The absolute value of $\theta_{LD}$ is in agreement with that of the
VLTI measurement in \citet{2005A&A...436..253T}, who quote
$\theta_{LD}=2.200\pm0.027\pm0.016$ mas (``statistical'' and
``systematic'' errors are cited separately). This previous
measurement is based upon limited data (only 3 $V^2$ data points)
and is anchored to the angular size estimates of
\citet{1999AJ....117.1864C}. Tracing back through the calibration
history outlined in their paper, this is an indication that the SED
angular size estimates from \citet{1999AJ....117.1864C} of resolved
calibrators $\alpha$ Crt and $\mu$ Vir they used to calibrate the
system were accurate at the stated uncertainties, although this is
uncertain -- no values were quoted in the manuscript -- and that
Th{\'e}venin et al's measurement process preserved that SED
accuracy.  Additionally, no evaluation of the impact of possible
binarity was considered by \citet{2005A&A...436..253T}, possibly due
to the limits of their relatively small sample. Our result, since it
is anchored to unresolved calibration sources, avoids the danger of
being susceptible to any potential significant systematic error from
SED modeling.

\subsection{$T_{EFF}$ and $R$}

The effective temperature can be
directly derived from the bolometric flux and the limb-darkening angular size:
\begin{equation}
T_{EFF}=2341 \times \left[ {F_{BOL} \over \theta^2_{LD}}
\right]^{1/4}
\end{equation}
where $F_{BOL}$ are in $10^{-8}$ ergs cm$^{-2}$ s$^{-1}$ and $\theta_{LD}$ is in mas
\citep{1999AJ....117..521V}.
Stellar radius is given by $R=0.1076 \theta_{LD} d$; where $R$ is in
$R_\odot$, $d$ is in parsecs, and $\theta_{LD}$ is used as a proxy for
the Rosseland mean diameter.
Luminosity can be derived directly from the radius and effective temperature,
$L=4 \pi R^2 \sigma T_{EFF}^4$
and is wholly independent of {\it PTI}'s measure of $\theta$, being depending only on $d$ and
our estimate of $F_{BOL}$ (and, by extension, $A_V$).
These derived values are summarized in Table \ref{table_starSummary}.
Our value of $L=8.89\pm0.16 L_\odot$ is statistically consistent
with the independent \citet{2005ApJ...635..547G} value, indicating
our $F_{BOL}$ value discussed in \S \ref{sec_SED_fitting} is accurate.

The measured effective temperature for HD 121860, $T_{EFF}=3627 \pm
46 $ K, which differs only slightly from the expected $T_{EFF}=3750 \pm 22$ K
\citep{1999AJ....117..521V} for a M2III
spectral type
derived from the described fitting
in \S \ref{sec_SED_fitting}.
Its radius of $ 147 \pm 92 $ $R_\odot$
exceeds the expected value of $\sim 60 R_\odot$, but the
error on the radius is sufficently large (due to the large error on the parallax)
that it is not inconsistent with the smaller expected
radius.
The agreement of HD121860's $T_{EFF}$ with the expected value was
an indication to us that our confidence in {\it PTI}'s data and its error
estimates are reasonable.

\section{Non-Detection of  Binarity of $\eta$ Boo}\label{sec_possibleBinary}

$\eta$ Boo has historically been listed as a possible spectroscopic
binary \citep{1957ApJ...125..696B,1982Ap&SS..87..377V}, with an
orbital confirmation being cited by \citet{1976ApJS...30..273A}.
However, speckle interferometery observations
\citep{1978PASP...90..288M,1980A&AS...42..185B,1984PASP...96..105H,1992AJ....104.1961M}
showed no evidence for detection of a possible companion to $\eta$
Boo, despite an expected angular separation of $0.170"$
\citep{1981A&AS...44...47H}, and particularly given an quoted
luminosity ratio of $\approx 0.75$ \citep{1982Ap&SS..87..377V}.
However, the original discovery paper for $\eta$ Boo's binary nature
\citep{1957ApJ...125..696B} suggests a late K- or M-type dwarf
companion to the G0IV subgiant primary, which would indicate a
brightness difference of at least $\Delta K \approx 4.5$, a
substantially greater value than indicated by
\citet{1982Ap&SS..87..377V}, and one consistent with the speckle
non-detections. \citet{1976ApJS...30..273A} also cite an earlier
astrometric detection \citep{dan39}, although this seems unlikely
given both the expected separation and the brightness ratio.
The spectroscopy of \citet{1997ApJ...475..322B} indicates
$\eta$ Boo's barycentric velocity
being influenced by binarity, which \citet{1995ApJ...443L..29C} have
suggested is indeed due to a M dwarf.

Nevertheless, given the higher resolution of {\it PTI}, a possible
close-separation secondary companion may affect our measures of
$\eta$ Boo's $V^2$ and thereby complicate our interpretation.  As such,
it was prudent for us to examine our data for evidence of $\Delta
V^2$ excursions indicative of binarity. As seen in Figure
\ref{fig_etaBooAbsV2} and the data contained in Table \ref{table1},
the $\eta$ Boo $V^2$ data are consistent with a single star
hypothesis, incorporating a wide range of $(u,v)$ coverage and
dates.

To explore to what extent a secondary star could be hidden within our data points,
we examined in detail the residuals found in our single star fit,
as seen in the bottom panel of Figure \ref{fig_etaBooAbsV2}.
We began by creating a synthetic $V^2$ dataset corresponding to a single
star observed by PTI with $\theta_{LD}$ = 2.2 mas, to which we added
varying amounts of measurement noise.  We then characterized the
$V^2$ residuals through use of a histogram created through
averaging two dozen runs of this synthesis, stepping through the residual
values in increments of $\Delta \sigma_{V^2} = 0.01$, and comparing that
histogram to that of our actual data.  We found that the best fit was
for measurement noise at the $\sigma_{V^2} = 0.018$ level with reduced $\chi^2$ fit
value of $\chi_\nu^2=1.66$, consistent with
our expectation for PTI data discussed in \S \ref{sec_SED_fitting}.

We then repeated this process, including measurement noise at
the $\sigma_{V^2} = 0.018$ level but also adding a main sequence stellar companion
of varying spectral type, using standard values of luminosity
and color from \citet{cox00}, and constraining the orbital parameters
to match the reported period of $P=494^d$.  The reduced $\chi^2$ fit
value increased as expected as the mass of the putative companion increased,
but we could not exclude a possible companion of spectral type M7 and lower in a
statistically meaningful way.  The net result of any potential $V^2$ bias of
a M7 companion with $\Delta K \approx 5.5$ would be to decrease the actual size of the $\eta$ Boo primary
by 0.014 mas.  As such, we are including an additional, negative error in our calculations
for the absolute parameters of $\eta$ Boo to account for our uncertainty
regarding this possible companion.

Given the long
time baseline of $V^2$ measures that are present in our full
dataset, it seems unlikely that chance geometric
alignment would persist in making a secondary companion not appear
in the form of $\Delta V^2$ excursions. It is still entirely
possible that a secondary companion is present, but at a brightness
ratio that makes it not a factor in determining $\eta$ Boo's size
($L_2 / L_1 < 0.005$), which is consistent with the original
\citet{1957ApJ...125..696B} result.

\begin{deluxetable}{lccccccc}
\tablecolumns{8}
\tablewidth{0pc}
\tablecaption{Summary of stellar fundamental parameters measured for
$\eta$ Boo and HD121860.  The error on $\theta_{LD}$ includes the uncertainty
discussed in \S \ref{sec_possibleBinary}; subsequent columns propagate
the larger of the two error bars where appropriate.\label{table_starSummary}}
\tablehead{
\colhead{Star} & \colhead{$\theta_{LD}$} & \colhead{$F_{BOL}$} & \colhead{$\pi$\tablenotemark{a}} &  \colhead{$T_{EFF}$} & \colhead{$R$} & \colhead{$L$}& \colhead{$\log g$}\\
               & \colhead{(mas)}         & \colhead{($10^{-8}$ erg cm$^{-2}$s$^{-1}$)}
                                                               & \colhead{(mas)}   &  \colhead{(K)} & \colhead{($R_\odot$)} & \colhead{($L_\odot$)} & \colhead{($\log$[cm s$^{-2}$])}
} \startdata


   $\eta$ Boo & $ 2.1894^{+0.0055}_{-0.0140} $ & $ 221 \pm 2.82   $ & $ 88.17 \pm 0.75 $ & $ 6100 \pm 28 $ & $ 2.672 \pm 0.028 $ & $ 8.89 \pm 0.16 $ & $ 3.817 \pm 0.016 $ \\

  HD121860 & $ 2.100 \pm 0.009 $ & $ 25.4 \pm 1.26   $ & $ 1.54 \pm 0.97 $ & $ 3627 \pm 46 $ & $ 147 \pm 92 $ & $ 3350 \pm 2989 $ & $ 0.184 \pm 0.288 $ \\

\enddata
\tablenotetext{a}{\citet{1997A&A...323L..49P}}
\end{deluxetable}



\section{Discussion the Astrophysical Parameters of  $\eta$ Boo}

\begin{figure*}
\plotone{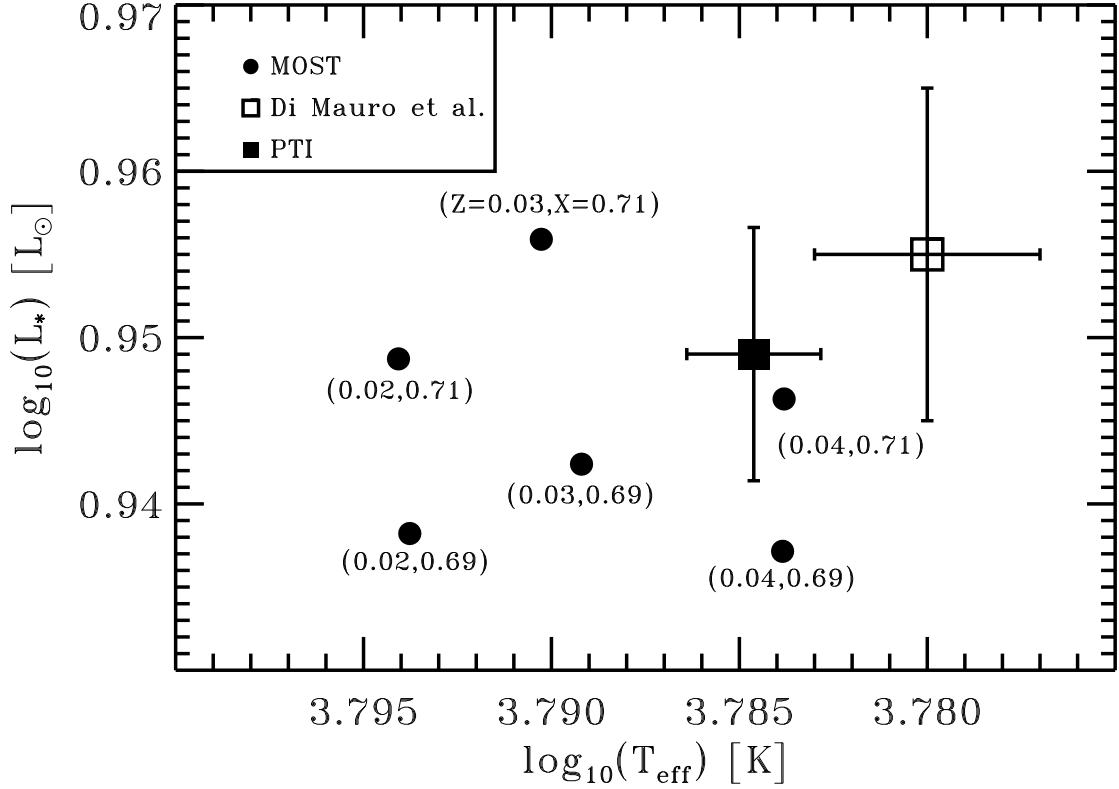}
\caption{\label{fig_fig2} Comparison of {\it MOST} fits to luminosity and effective temperature
values determined in this study and by \citet{dim03}.  The {\it MOST} fit corresponding to a model
composition of $\{X=0.71, Z=0.04\}$ was within
the error ellipse of this study's $\{\log L =
0.9490 \pm 0.0076,\log T = 3.7853 \pm 0.0020\}$ coordinates, and was ultimately
favored in \citet{2005ApJ...635..547G} as the best fit to their data.}
\end{figure*}

Placing $\eta$ Boo on a H-R diagram, just as was done in
\citet{2005ApJ...635..547G} (see their Figure 6), we may compare our
values for the star's luminosity and effective temperature to those
best fit to the {\it MOST} data.
We have highlighted this region of
interest in Figure \ref{fig_fig2}. We find our $\{\log L =
0.9490 \pm 0.0076,\log T = 3.7853 \pm 0.0020\}$ coordinates fall
within the locus of points defined by the models the {\it MOST}
team.
The error ellipse
of our $\{\log L,\log T\}$
derived from the PTI data
encompasses the {\it MOST}
$\{X=0.71, Z=0.04\}$ composition model point, which ultimately was favored by
Guenther et al. as the best fit to their {\it MOST} data.

The $\{\log L,\log T\}$ coordinates defined by \citet{dim03} were
sufficiently displaced from all of the possible {\it MOST}
coordinates that Guenther et al. suggested the \citet{dim03}
coordinates were incorrect.  Our $\eta$ Boo data and analysis are
clearly consistent with this suggestion by
\citet{2005ApJ...635..547G}.

\subsection{Surface Gravity of $\eta$ Boo}
With the mass for $\eta$ Boo established from the asteroseismic constraints
of {\it MOST}, and the radius from interferometric observations, we may directly
establish the surface gravity for $\eta$ Boo:
\begin{equation}
g = {G M \over R^2}
\end{equation}
Using the {\it MOST} derived mass of $M=1.71 \pm 0.05 M_\odot$ and
the {\it PTI} derived radius of $R=2.672 \pm 0.024 R_\odot$, we
derive a surface gravity of $\log g = 3.817 \pm 0.015$ [cm s$^{-2}$].

The {\it PTI} $\eta$ Boo $\log g$ result is in significant
disagreement with the ``trigonometric'' gravities found in
\citet{1999ApJ...527..879A}, who found $\log g = 3.47 \pm 0.10$ [cm
s$^{-2}$].  \citet{1990A&AS...85.1015M} quote a value of $\log g=3.8
\pm 0.15$ [cm s$^{-2}$] from stellar evolution theories, although
their angular size, radius and mass values on which those theories
were baselined are divergent from the {\it PTI} and {\it MOST}
results at 10-20\% level. The study by \citet{1985ApJS...57..389L}
using photometric and spectrophotometric data for $T_{EFF}$, and
Str\"omgren photometry plus intermediate dispersion spectra for
$\log g$, is in reasonable agreement with our values, quoting
$T_{EFF}= 5930 \pm 70$ K and $\log g = 3.71 \pm 0.15$ [cm s$^{-2}$].
Similarly, \citet{2002ApJ...577..377M} find $\log g = 3.71 \pm 0.13$
[cm s$^{-2}$] using ultraviolet-visual spectrophotometry. In all of
these cases, indirect measures of $\log g$ and other astrophysical
parameters are overshadowed by the more direct, empirical methods
afforded by the unique capabilities of {\it PTI} and {\it MOST}.

\section{Conclusions}

While considerably more accurate, our angular diameter determination
is statistically consistent with the earlier measurement from
\citet{2005A&A...436..253T}, an apparent indication that these
results are free from systematic error at their stated level on
uncertainty. Our angular diameter and bolometric flux values have
led to an effective temperature and luminosity for the evolved star
$\eta$ Boo that is in direct agreement with those established by the
{\it MOST} asteroseismology mission. Furthermore, in conjunction
with the {\it MOST} value for stellar mass, a measure of stellar
surface gravity may be made.

The combination of the precise mass estimate from {\it MOST} and the
accurate radius measure of {\it PTI} has allowed us to derive a
precise value of the surface gravity for $\eta$~Boo. The measurement
of the surface gravity $\eta$ Boo, independent of spectroscopy, is a
significant demonstration of the astrophysical investigative value
of combining high spatial resolution interferometry with high
temporal resolution photometry.  Of the determinations of $\log g$
for $\eta$ Boo in the literature, we find that our value has a
claimed precision an order of magnitude greater than previous
measures.

\acknowledgements

We gratefully acknowledge fruitful discussions with Jaymie Matthews
and Theo ten Brummelaar. Science operations with PTI are conducted
through the efforts of the PTI Collaboration
(http://pti.jpl.nasa.gov/ptimembers.html), and we acknowledge the
invaluable contributions of our PTI colleagues.  We particularly
thank Kevin Rykoski for his professional operation of PTI. This
research has made use of the SIMBAD database, operated at CDS,
Strasbourg, France. Funding for PTI was provided to the Jet
Propulsion Laboratory under its TOPS (Towards Other Planetary
Systems), ASEPS (Astronomical Studies of Extrasolar Planetary
Systems), and Origins programs and from the JPL Director's
Discretionary Fund. Portions of this work were performed at the Jet
Propulsion Laboratory, California Institute of Technology under
contract with the National Aeronautics and Space Administration.

\end{document}